\documentclass{emulateapj}

\usepackage{amssymb, amsmath, apjfonts, natbib, color}

\def\arcsec{{\prime\prime}}
\def\arcmin{{\prime}}
\def\degree{{\circ}}
\newdimen\digitwidth
\setbox1=\hbox{0}
\digitwidth=\wd1
\catcode`"=\active
\def"{\kern\digitwidth}

\slugcomment{Accepted for publication in The Astrophysical Journal}

\begin{document}

\title{The Carnegie-Irvine Galaxy Survey. V. Statistical study of bars and buckled bars}

\author{Zhao-Yu Li\altaffilmark{1, 2, 6}, Luis C. Ho\altaffilmark{3, 4}, Aaron J. Barth\altaffilmark{5}}

\altaffiltext{1}{Key Laboratory for Research in Galaxies and Cosmology, Shanghai Astronomical Observatory, Chinese Academy of Science, 80 Nandan Road, Shanghai 200030, China; lizy@shao.ac.cn}
\altaffiltext{2}{College of Astronomy and Space Sciences, University of Chinese Academy of Sciences, 19A Yuquan Road, Beijing 100049, China}
\altaffiltext{3}{Kavli Institute for Astronomy and Astrophysics, Peking University, Beijing 100871, China}
\altaffiltext{4}{Department of Astronomy, School of Physics, Peking University, Beijing 100871, China}
\altaffiltext{5}{Department of Physics and Astronomy, 4129 Frederick Reines Hall, University of California, Irvine, CA, 92697-4575, USA}
\altaffiltext{6}{Lamost Fellow}

\begin{abstract}

Simulations have shown that bars are subject to a vertical buckling instability that transforms thin bars into boxy or peanut-shaped structures, but the physical conditions necessary for buckling to occur are not fully understood. We use the large sample of local disk galaxies in the Carnegie-Irvine Galaxy Survey to examine the incidence of bars and buckled bars across the Hubble sequence. Depending on the disk inclination angle ($i$), a buckled bar reveals itself as either a boxy/peanut-shaped bulge (at high $i$) or as a barlens structure (at low $i$). We visually identify bars, boxy/peanut-shaped bulges, and barlenses, and examine the dependence of bar and buckled bar fractions on host galaxy properties, including Hubble type, stellar mass, color, and gas mass fraction. We find that the barred and unbarred disks show similar distributions in these physical parameters. The bar fraction is higher (70\%--80\%) in late-type disks with low stellar mass ($M_{*} < 10^{10.5}\, M_{\odot}$) and high gas mass ratio. In contrast, the buckled bar fraction increases to 80\% toward massive and early-type disks ($M_{*} > 10^{10.5}\, M_{\odot}$), and decreases with higher gas mass ratio. These results suggest that bars are more difficult to grow in massive disks that are dynamically hotter than low-mass disks. However, once a bar forms, it can easily buckle in the massive disks, where a deeper potential can sustain the vertical resonant orbits. We also find a probable buckling bar candidate (ESO 506$-$G004) that could provide further clues to understand the timescale of the buckling process.

\end{abstract}


\section{INTRODUCTION}
\label{sec:intro}

Bars are common features in disk galaxies, and they play important roles in secular evolutionary processes \citep{kor_ken_04}. Optical studies indicate that the bar fraction is $\sim$50\% \citep{mar_jog_07, barazz_etal_08, aguerr_etal_09}, while near-infrared surveys find even higher bar fractions of $\sim$70\% \citep{hac_sch_83, eskrid_etal_00, eskrid_etal_02, jarret_etal_03, menend_etal_07}. The existence of a bar in the central region of a galaxy makes it an important probe of the dynamics of the inner disk \citep{aguerr_etal_09}. Bar structure depends on the overall distributions of mass and angular momentum in the host galaxy, and the dynamical effect of a bar on both baryonic and dark matter affects the structural assembly history and morphological evolution of the galaxy \citep{weinbe_85, deb_sel_98}. The bar can effectively funnel gas toward the galaxy center to induce the build-up of the central stellar concentration or pseudobulge \citep{com_ger_85, athana_92b, reg_elm_97, sakamo_etal_99, sheth_etal_00, kor_ken_04}. This inflow can lead to dramatic changes in the galaxy, such as the smoothing of chemical abundant gradients \citep{mar_roy_94}, formation of nuclear star-forming rings \citep{buta_etal_03} or nuclear bars \citep{erwin_04}, and fueling of the central black hole \citep{shlosm_etal_89}. 

Numerical simulations have shown that a bar forms via the bar instability, and then thickens vertically due to the buckling instability \citep{com_san_81, raha_etal_91}. The bar quickly buckles in the vertical direction, resulting in a boxy/peanut-shaped (B/PS) bulge in the inner part of the bar  as seen in an edge-on view. In extreme cases the inner region may develop a pronounced X-shaped structure \citep{li_she_12}. The three-dimensional isodensity surfaces of numerically simulated buckled bars are actually composed of three components: the central boxy core, the intermediate B/PS bulge, and an extended thin bar \citep{li_she_15}. Such simulations can trace the process of bar formation and the subsequent buckling. However, it is not trivial to test models or simulations of bar evolution with observations. Finding buckled bars in real galaxies is challenging, because of projection effects: a buckled bar shows boxy/peanut-like structure only in an edge-on view. At smaller inclinations, the morphological characteristics of a buckled bar become more difficult to identify. 

\citet{erw_deb_13} have shown that at relatively moderate inclinations, a B/PS bulge can still be recognized by the orientation differences between the inner boxy isophotes and the outer spurs, which resemble narrow extensions to the isophotes at the outer part of the bar. They pointed out that the spurs are always shifted away from the major axis of the boxy inner zone; this originates from the result of viewing a bar with an inner thickened boxy/peanut bulge and an extended thin component. The B/PS bulge creates the boxy isophotes, while the outer flat part of the bar forms the spurs. Moreover, the outer spurs are offset in the opposite directions from the major axis of the inner boxy structure. 

In face-on galaxies, the identification of a B/PS bulge is not straightforward due to the absence of boxy isophotes \citep{li_she_15}. Recently, a new structure, namely the ``barlens'' structure, has been identified in disk galaxies in the Near-IR S0 galaxy Survey (NIRS0S; Laurikainen et al. 2011).  It features a lens-like morphology embedded in the bar, covering typically half of the bar length. The outline of the barlens is usually oval or circular, with much smaller ellipticity compared to the extended bar. Both simulations and observations suggest that the barlens is, in fact, the face-on view of a B/PS bulge in the inner half of the bar \citep{athana_etal_15, lau_sal_17}. 

Although bar formation and growth have been studied extensively in numerical simulations, direct evidence of bar evolution is still lacking from the observational point of view. The identification of B/PS bulge and barlens structures provides an opportunity to investigate important properties related to bar formation and evolution, such as the galaxy stellar mass and the gas mass fraction. Although simulations can easily generate bars and B/PS bulges in disks, the necessary conditions for bars to form and buckle are still not well understood from observations. To answer these questions, a large and statistically complete sample of disk galaxies is essential.

Recently, \citet{erw_deb_17} selected a sample of 84 local barred, moderately inclined disk galaxies to estimate the fraction of B/PS bulges. They found that the frequency of B/PS bulges strongly depends on the stellar mass, with higher fraction of B/PS bulges in massive galaxies ($> 10^{10.4}\ M_{\odot}$). They also found a high B/PS fraction for S0-Sbc galaxies ($\sim$80\%), and lower values for later-type disk galaxies (15\%). 

Here, we use the Carnegie-Irvine Galaxy Survey (CGS) as an independent large sample to study the statistical incidence of bars and buckled bars. CGS is a complete broad-band photometric survey of nearby galaxies in the southern hemisphere. As described by Ho et al. (2011; hereafter Paper I) and Li et al. (2011; hereafter Paper II), the CGS images have very good signal-to-noise ratio (S/N) and high resolution. Compared to \citet{erw_deb_17}, our sample is larger and more statistically complete. In addition, we identify both B/PS bulges and barlens structures in face-on and moderately inclined disks simultaneously. This can give us a much more comprehensive picture about bars and buckled bars.

In Section 2, we describe the sample selection and identification of buckled bars. The results are shown in Section 3. In Section 4, we discuss the implications of the results, and compare with previous work. Our results are summarized in Section 5.

\begin{figure}
\epsscale{1.2}
\plotone{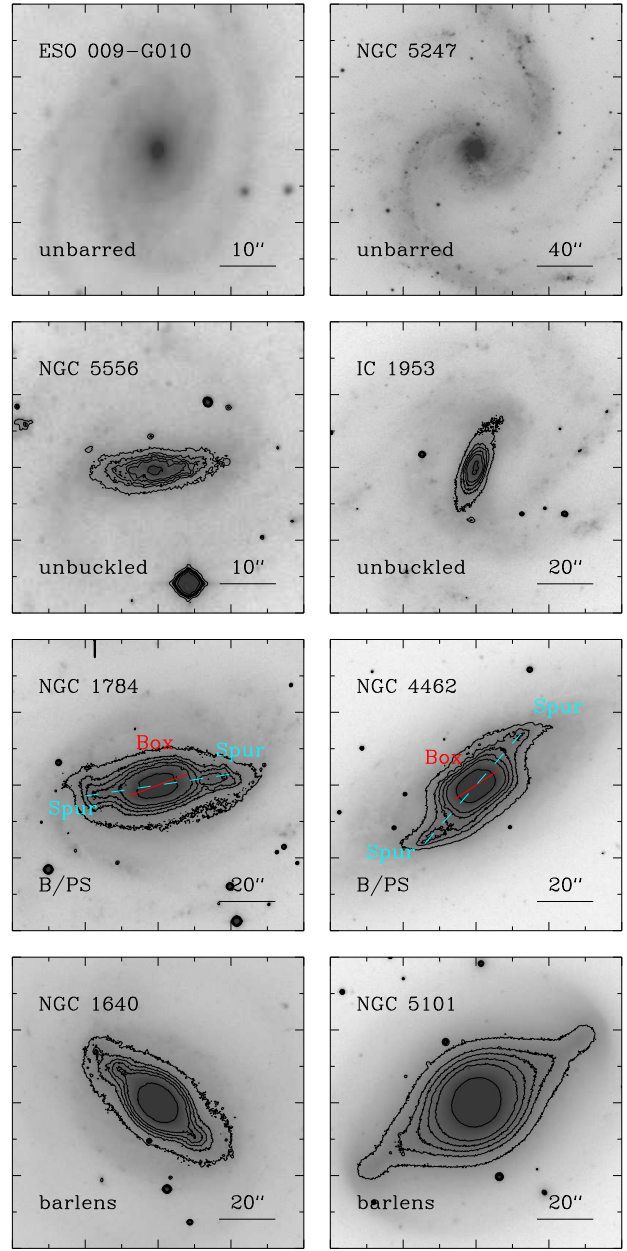}
\caption{Examples of unbarred disks (top row), unbuckled bars (second row), B/PS bulges (third row) and barlenses (bottom row) from CGS $I$-band images, with North up and East to the left. In the third and fourth rows, the isodensity contours are overlaid to visualize the buckled bar. The isodensity contours of the unbuckled bars in the second row are also shown for comparison. For the B/PS bulge in the third row, the red solid line represents the major axis of the inner boxy structure, and the cyan dashed line represents the direction of the spurs. The spurs are offset from the major axis of the boxy bulge. For the barlens, the outlines of the isophotes show smaller ellipticity as a lens, which traces the vertically thickened part of the bar.}
\epsscale{1.}
\label{fig:demo}
\end{figure}

\begin{figure}
\epsscale{1.2}
\plotone{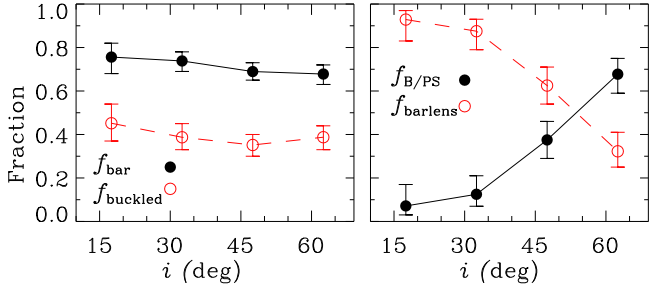}
\caption{The bar fraction ($f_{\rm bar} = N_{\rm bar} / N_{\rm disk}$), buckled bar fraction ($f_{\rm buckled} = N_{\rm buckled} / N_{\rm bar}$), B/PS fraction ($f_{\rm B/PS} = N_{\rm B/PS} / N_{\rm buckled}$), and barlens fraction ($f_{\rm barlens} = N_{\rm barlens} / N_{\rm buckled}$) at different inclination angles.}
\label{fig:frac_inc}
\epsscale{1.}
\end{figure}

\begin{figure*}
\plotone{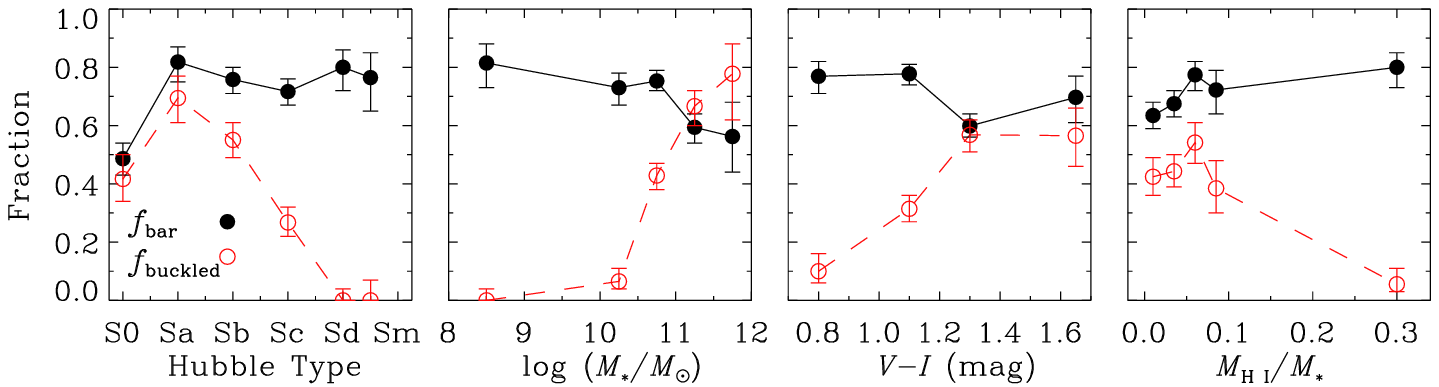}
\caption{Bar fraction in disk galaxies ($f_{\rm bar} = N_{\rm bar} / N_{\rm disk}$; black solid line) and buckled bar fraction in barred galaxies ($f_{\rm buckled} = N_{\rm buckled} / N_{\rm bar}$; red dashed line) as a function of the Hubble type (first panel), galaxy stellar mass (second panel), color (third panel), and gas mass fraction (fourth panel).}
\label{fig:frac}
\end{figure*}


\section{SAMPLE AND METHOD}
\label{sec:samp}

\subsection{Sample selection}

CGS contains a statistically complete sample of 605 bright, nearby galaxies of all morphological types in the southern hemisphere, with $B$-band total magnitude $B_T \leq 12.9$ and $\delta < 0^\degree$. Given these general selection criteria, the sample includes a broad distribution of galaxy types and luminosities. Images in $B, V, R$ and $I$-band filters were taken with the du Pont 2.5-m telescope at Las Campanas Observatory, with a field-of-view (FOV) of $8.^\arcmin9 \times 8.^\arcmin9$ and pixel scale of $0.^\arcsec259$. 
The seeing is about $1^\arcsec$ over all bands. In the $I$ band, seeing better than $1^\arcsec$ was often obtained. The typical depths of the $B$, $V$, $R$ and $I$-band images are 27.5, 26.9, 26.4, and 25.3 mag $\rm arcsec^{-2}$, respectively (defined as the surface brightness of outer galaxy isophotes that are $1\sigma$ above the background sky rms). Thus, for studies of galaxy morphology, CGS provides the advantages of high S/N, high spatial resolution, and a relatively large sample size. More information about the survey design, data reduction, and photometric measurements can be found in Papers I and II. 

For this study, we select a disk galaxy sample based on the following criteria. Starting with the CGS sample, we first exclude ellipticals and interacting galaxies. We restrict the sample to galaxies with inclination angle ($i$) less than $70^\degree$. Each galaxy is classified as barred or unbarred according to the morphological identification in the Third Reference Catalogue of Bright Galaxies (RC3; \citealt{devauc_etal_91}). We have visually examined all the $I$-band images of the sample and found good agreement with the RC3 morphological classification\footnote[1]{Our visual examination indicates that three previously identified unbarred galaxies (NGC 3281, NGC 3885, and NGC 7079) should be classified as weakly barred.}. The sample contains 376 disk galaxies with 264 disks hosting bars. In this work, we use the CGS $I$-band images to minimize the effect of dust extinction. This is important because bar kinematics can channel gas into strong dust lanes in the leading side of the bar. These features can bias the isophotal contours in the bar region to mislead our classification in the optical bands.

\begin{deluxetable*}{lcccccr} 
\tabletypesize{\footnotesize} 
\tablecolumns{7} 
\tablewidth{0pt} 
\tablecaption{CGS Disk Galaxy Properties and Bar Classification} 
\tablehead{
\colhead{Name} & \colhead{Morphology} & \colhead{$i$}& \colhead{$\log{M_{*}}$} & \colhead{$V-I$} & \colhead{$M_{\rm H~I} / M_{*}$} & \colhead{Bar Type} \\
\colhead{} & \colhead{} & \colhead{($^\degree$)} & \colhead{($M_{\odot}$)} & \colhead{(mag)} & \colhead{} & \colhead{} \\
\colhead{(1)} & \colhead{(2)} & \colhead{(3)} & \colhead{(4)} & \colhead{(5)} & \colhead{(6)} & \colhead{(7)}}
\startdata 

 ES0 009$-$G010 &    Sbc & $45.63$ & $10.79$ & $ 1.22$ & $ 0.01$ &          unbarred\\
 ES0 027$-$G001 &    SBc & $45.01$ & $10.22$ & $ 0.90$ & $ 0.14$ &         unbuckled\\
 ES0 060$-$G019 &   SBcd & $46.03$ & $ 9.80$ & $ 1.08$ & $ 0.19$ &         unbuckled\\
 ES0 091$-$G003 &    Sab & $53.34$ & $10.54$ & $ 1.04$ & $ 0.01$ &          unbarred\\
 ES0 121$-$G026 &   SBbc & $44.94$ & $10.98$ & $ 1.15$ & $ 0.03$ &           barlens\\
 ES0 136$-$G012 &    SBc & $32.76$ & $10.33$ & $ 1.12$ & $ 0.28$ &         unbuckled\\
 ES0 137$-$G034 &   S0/a & $38.39$ & $11.09$ & $ 1.49$ & $ 0.01$ &         unbuckled\\
 ES0 138$-$G010 &     Sd & $47.18$ & $10.42$ & $ 1.36$ & $ 0.09$ &          unbarred\\
 ES0 138$-$G029 &   S0/a & $51.76$ & $11.46$ & $ 0.00$ & $ 0.00$ &         unbuckled\\
 ES0 183$-$G030 &   E/S0 & $40.38$ & $11.04$ & $ 1.36$ & $ 0.01$ &          unbarred\\
 ES0 186$-$G062 &   SBcd & $28.38$ & $10.29$ & $ 1.07$ & $ 0.12$ &         unbuckled\\
 ES0 208$-$G021 &   E/S0 & $44.49$ & $10.67$ & $ 1.32$ & $ 0.02$ &         unbuckled\\
 ES0 213$-$G011 &     Sc & $59.29$ & $10.79$ & $ 1.19$ & $ 0.03$ &          unbarred\\
 ES0 221$-$G032 &     Sc & $42.85$ & $10.94$ & $ 1.23$ & $ 0.02$ &          unbarred\\
 ES0 269$-$G057 &   SABa & $40.35$ & $11.18$ & $ 1.29$ & $ 0.06$ &           barlens\\
 ES0 271$-$G010 &     Sc & $36.06$ & $10.47$ & $ 1.04$ & $ 0.02$ &         unbuckled\\
 ES0 320$-$G026 &     Sb & $69.73$ & $11.18$ & $ 1.33$ & $ 0.01$ &          unbarred\\
 ES0 321$-$G025 &    SBc & $69.87$ & $10.47$ & $ 1.02$ & $ 0.04$ &         unbuckled\\
 ES0 380$-$G001 &     Sb & $62.02$ & $10.74$ & $ 1.30$ & $ 0.05$ &              B/PS\\
 ES0 380$-$G006 &     Sb & $64.09$ & $11.21$ & $ 1.33$ & $ 0.01$ &          unbarred\\
 ES0 383$-$G087 &     Sd & $43.88$ & $ 8.27$ & $ 1.00$ & $ 0.44$ &         unbuckled\\

\enddata 
\tablecomments{Col. (1): Galaxy name. Col. (2): Galaxy morphology. Col. (3): Galaxy inclination angle measured in the $I$ band. Col. (4): Stellar mass. Col. (5): Extinction-corrected $V-I$ color of the whole galaxy. Col. (6): H~I mass ratio. Col. (7): Bar classification. {\it Table~\ref{tab:class} is presented in its entirety in the electronic edition of the journal. A portion is shown here for guidance regarding its form and content.}} 
\label{tab:class}
\end{deluxetable*}

\subsection{Buckled bar identification}

A buckled bar shows B/PS isophotes at large inclinations. \citet{erw_deb_13} have shown that direct detection of a B/PS bulge is possible not only for edge-on galaxies, but also for galaxies with moderate inclination angles ($i < 70^\degree$). A clear example is given in Figure~1 of \citet{erw_deb_13}. The inner part of the bar has ``boxy'' isophotes, while the outer isophotes of the bar appear narrower at the end of the boxy zone. These features are denoted as ``spurs''. The spurs are offset from the major axis of the inner boxy structure in opposite directions. In addition, they also suggested that elliptical isophotal analysis  may not be suitable to detect such features, because the boxy+spur isophotes are not well fit by ellipses. In fact, they recommended to measure the radius of the boxy region directly from the images.

The B/PS structure is not directly apparent in face-on disk galaxies. \citet{li_she_15} have shown that the face-on projection of an inner vertically thickened B/PS bulge has nearly elliptical isophotal structure. But it is still possible to trace the B/PS bulge in this case. \citet{athana_etal_15} examined the properties of barlenses, which are lens-like components inside bars whose outline is oval or circular. By comparing numerical simulations from different viewing angles with observations, they concluded that a barlens is the vertically thick portion of the bar seen face-on. For identification of barlenses, they suggested that any automatic ellipse fitting would not be useful, because the fit would be a compromise between the shape of the barlens and that of the thin outer parts of the bar, which would not give an accurate description of the barlens isophotes. 

Using the {\it Spitzer}\ Survey of Stellar Structure in Galaxies (S$^4$G) \citep{sheth_etal_10, munozm_etal_13, querej_etal_15} and NIRS0S, \citet{lau_sal_17} identified barlenses in galaxies spanning a large range of inclinations ($i \leq 60^\degree$). Comparisons with $N$-body simulations led them to conclude that the ``observed'' barlens feature can be a result of both the galaxy inclination and the central mass concentration. They suggested that the barlens morphology is expected at $i = 50^\degree$ when at least a few percent of the disk mass is in a central component; the inner bulge can enhance the appearance of the barlens structure at relatively larger inclination angles.

For the 264 barred galaxies in our CGS sample, we visually search for and identify buckled bar characteristics, namely B/PS bulges and barlenses, according to the criteria mentioned above \citep{laurik_etal_11, erw_deb_13, athana_etal_15, lau_sal_17}. Starting from the $I$-band images of the barred galaxies, we first extract the isophotal contours in the bar region. Our visual examination is mainly based on the contours inside the bar. If the contour looks boxy with a pair of offset spurs connecting at the end, we classify the bar as a B/PS bulge. If the contour is oval or circular with smaller ellipticity compared to the outer contours of the bar, it is classified as a barlens. Of course, even in the $I$-band, there are some cases with severe dust extinction in the bar region, producing distorted isophotal contours. In this situation, we rely on lower resolution S$^4$G and 2MASS \citep{skruts_etal_06} images to make the final decision.

In this sample, we find 101 buckled bar candidates, with 37 B/PS bulges and 64 barlenses. From top to bottom rows, Figure~\ref{fig:demo} shows examples of unbarred disks, unbuckled bars, B/PS bulges, and barlenses. In the third and fourth rows, isophotal contours are overlaid to aid visualizations of B/PS bulges and barlenses. For the B/PS bulges, there is a clear offset between the direction of the spurs (cyan dashed line) and the direction of the inner boxy isophotes (red solid line) \citep{erw_deb_13}. For the barlenses, the inner isophotes generally have smaller ellipticities to represent a lens-like structure, which traces the vertically thickened part of the bar \citep{athana_etal_15}. Table~\ref{tab:class} lists the physical parameters and morphological classification of our sample. The stellar mass is estimated from the $K$-band magnitudes listed in Paper I. We adopt a single $K$-band stellar mass-to-light ratio of 0.75, which is the average value for the luminous galaxies studied by \citet{bell_etal_03} and adjusted to a \citet{chabri_03} initial mass function. We use the extinction-corrected value of $V - I$ to represent the integrated color of each galaxy. The gas mass fraction is derived from the neutral atomic hydrogen gas mass to $B$-band luminosity ratio ($M_{\rm H~I} / L_{B}$) and $B$-band luminosity listed in Paper I.  


\section{RESULTS}
\label{sec:results}

\subsection{Distribution of bar and buckled bar fractions}

In this section, we study the physical parameter differences between barred and unbarred disks, and between buckled bars and unbuckled bars. The main purpose is to understand the origin of bar formation and examine the conditions under which bars buckle. First, we check the bar and buckled bar fraction at different inclination angles. From the left panel of Figure~\ref{fig:frac_inc}, the bar fraction ($f_{\rm bar} = N_{\rm bar} / N_{\rm disk}$) and buckled bar fraction ($f_{\rm buckled} = N_{\rm buckled} / N_{\rm bar}$) are roughly constant across different inclination angles. This suggests that the RC3 bar identification is robust and not affected by the inclination angle. In the right panel of Figure~\ref{fig:frac_inc}, the B/PS fraction ($f_{\rm B/PS} = N_{\rm B/PS} / N_{\rm buckled}$) increases toward higher inclination angles, while the barlens fraction ($f_{\rm barlens} = N_{\rm barlens} / N_{\rm buckled}$) decreases significantly. This is expected from the 3-D morphology of the buckled bars, where the isophotes are more boxy at larger inclination angles, confirming that our visual identification results are probably unaffected by the inclination angles. 

At $60^\degree$ inclination, the barlens fraction in CGS is $\sim$30\%. This is consistent with previous studies. \citet{laurik_etal_14} performed a statistical study on the barlenses and boxy/peanut/X-shaped bulges in S$^4$G and NIRS0S. According to their Figure~2(a), the minor-to-major axis ratios ($b/a$) of galaxies hosting barlens have a wide distribution, from 0.3 to 1.0, roughly corresponding to $i = 70^\degree$ to face-on. Compared to the B/PS bulges, the relative fraction of barlenses increases from $\sim$ 30\% at $i \approx 60^\degree$ to 100\% at $i \approx 0^\degree$. Based on $N$-body simulations, \citet{lau_sal_17} concluded that the barlens feature can be a result of both the galaxy inclination and the central mass concentration. With a few percent disk mass within the central region much smaller than the barlens itself, the barlens feature can be detected at $i = 50^\degree$.


In Figure~\ref{fig:frac}, from left to right panels, we show the fraction of bars in disk galaxies ($f_{\rm bar} = N_{\rm bar} / N_{\rm disk}$) and the fraction of buckled bars in the bar sample ($f_{\rm buckled} = N_{\rm buckled} / N_{\rm bar}$) as a function of the galaxy morphology, stellar mass ($M_{*}$), color ($V-I$), and H~I gas mass fraction ($M_{\rm H~I} / M_{*}$), respectively. The profiles of Figures~\ref{fig:frac_inc} and~\ref{fig:frac} are also tabulated in Table~\ref{tab:stat}. The bar fraction in disk galaxies increases toward later types ($\sim$50\% in S0 galaxies to $\sim$70\% in Sd--Sm disks; first panel), lower stellar mass (from 50\% at $M_{*} \approx  10^{11.5}\ M_{\odot}$ to 80\% at $M_{*} \approx  10^{8.5}\ M_{\odot}$; second panel), bluer optical colors (60\% for $V-I \approx 1.6$ to 80\% for $V-I \approx 0.8$; third panel), and higher gas content (60\% for $M_{\rm H~I}/M_{*} \approx 0$ to 80\% for $M_{\rm H~I}/M_{*} \approx 0.3$; fourth panel). The latter trend is expected because the gas mass ratio is higher in late-type disks \citep{rob_hay_94}.

\begin{figure}
\epsscale{1.2}
\plotone{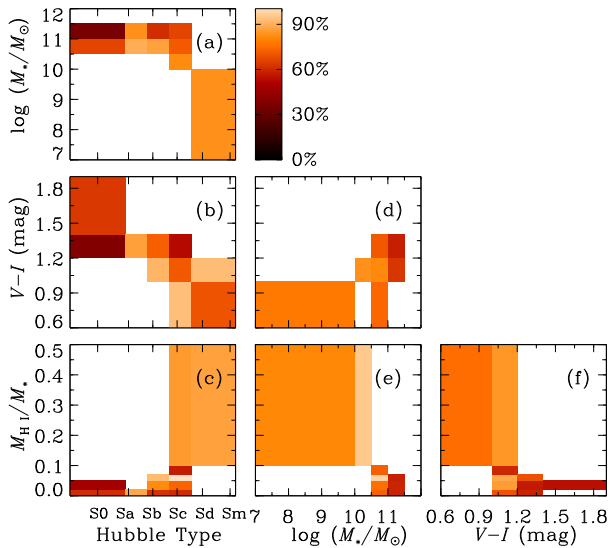}
\caption{The bar fraction ($f_{\rm bar} = N_{\rm bar} / N_{\rm disk}$) map in different physical parameter spaces.}
\label{fig:frac_grid_map}
\epsscale{1.}
\end{figure}

Galaxy properties are highly correlated themselves; for instance, massive disk galaxies are in general redder and of earlier type. A more physically meaningful approach is to study the distributions of $f_{\rm bar}$ and $f_{\rm buckled}$ in bins of different galaxy parameters. This can help avoid parameter correlations. In Figure~\ref{fig:frac_grid_map}, we show the dependence of bar fraction on galaxy morphology, stellar mass, color, and gas mass ratio. The grids do not have equal size. The purpose is to balance the number of galaxies in each grid, which is required to be larger than 20. In panel (a), in each galaxy morphological bin, the bar fraction is higher in lower mass galaxies. Interestingly, from panels (b) and (c), S0s seem to have relatively higher bar fraction in objects with redder color and lower gas mass ratio, whereas for the other types, the bar fraction is higher for bluer colors and higher gas mass ratios. From panels (d) and (e), the bar fraction is apparently much lower in massive disks. For a given stellar mass bin, $f_{\rm bar}$ varies slightly with other parameters. The results suggest that the stellar mass is a more important parameter than color and gas mass ratio.

\begin{figure}
\epsscale{1.2}
\plotone{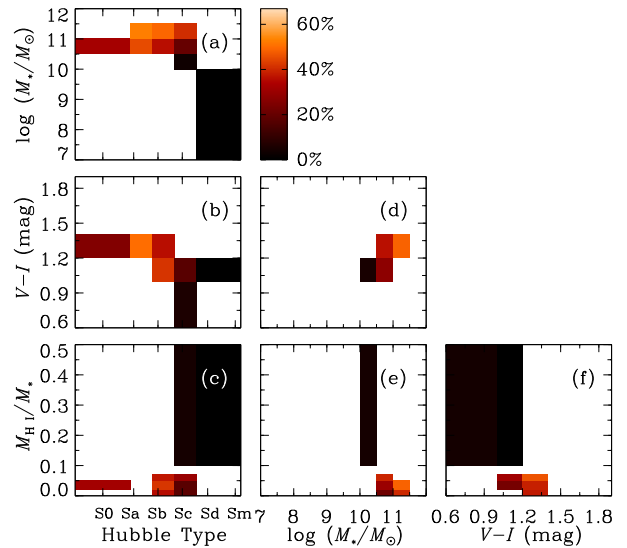}
\caption{The buckled bar fraction ($f_{\rm buckled} = N_{\rm buckled} / N_{\rm bar}$) map in different physical parameter spaces.}
\label{fig:frac_grid_buck}
\epsscale{1.}
\end{figure}

From the red dashed lines in Figure~\ref{fig:frac}, the buckled bar fraction shows much stronger variation with host galaxy parameters compared to the bar fraction. We find that $f_{\rm buckled}$ is almost 0 in galaxies with late-type morphology, lower stellar mass, blue color, and high gas mass ratios. On the other hand, the buckled bar fraction can reach 60\% in early-type morphologies, massive disks, red color, and lower gas mass ratios. Similar to the bar fraction, we also show the buckled bar fraction as a function of host galaxy properties in Figure~\ref{fig:frac_grid_buck}. We can see that $f_{\rm buckled}$ is highest in massive disks ($M_{*} > 10^{10.5} \, M_{\odot}$), in redder disks ($V-I >1.2$ mag), and in systems with lower gas mass ratios ($M_{\rm H~I} / M_{*} < 0.1$). This is also true in each morphological bin, as shown in panels (a)--(c). From panels (d) and (e), $f_{\rm buckled}$ = 0\% below $10^{10.5} \, M_{\odot}$, and reaches $\sim$ 60\% at $10^{11.5} \,M_{\odot}$. This result indicates that stellar mass is also the key parameter for the bar to buckle.

\begin{deluxetable*}{lcccccc} 
\tabletypesize{\footnotesize} 
\tablecolumns{7} 
\tablewidth{0pt} 
\tablecaption{Bar Fraction and Buckled Bar Fraction in Bins of Different Physical Parameters} 
\startdata 
\hline \hline
\\
 & $0^\degree < i < 25^\degree$ & $25^\degree < i < 40^\degree$ & $40^\degree < i < 55^\degree$ & $55^\degree < i < 70^\degree$ \\
\hline
\\
$N_{\rm disk}$ & 41 & 84 & 132 & 118 \\
$N_{\rm bar}$ & 31 & 62 & 91 & 80 \\
$N_{\rm buckled}$ & 14 & 24 & 32 & 31 \\
$N_{\rm B/PS}$ & 1 & 3 & 12 & 21 \\
$N_{\rm barlens}$ & 13 & 21 & 20 & 10 \\
$f_{\rm bar}$ & $0.76^{+0.06}_{-0.08}$ & $0.74^{+0.04}_{-0.05}$ & $0.69^{+0.04}_{-0.04}$ & $0.68^{+0.04}_{-0.05}$ \\
\\
$f_{\rm buckled}$ & $0.45^{+0.09}_{-0.08}$ & $0.39^{+0.06}_{-0.06}$ & $0.35^{+0.05}_{-0.05}$ & $0.39^{+0.05}_{-0.06}$ \\
\\
$f_{\rm B/PS}$ & $0.07^{+0.10}_{-0.04}$ & $0.13^{+0.09}_{-0.06}$ & $0.38^{+0.09}_{-0.09}$ & $0.68^{+0.07}_{-0.09}$ \\
\\
$f_{\rm barlens}$ & $0.93^{+0.04}_{-0.10}$ & $0.88^{+0.06}_{-0.09}$ & $0.63^{+0.09}_{-0.09}$ & $0.32^{+0.09}_{-0.07}$ \\
\\

\hline \hline
\\
 & S0 & Sa & Sb & Sc & Sd & Sm \\
\hline
\\

$N_{\rm disk}$ & 74 & 44 & 91 & 120 & 30 & 17 \\
$N_{\rm bar}$ & 36 & 36 & 69 & 86 & 24 & 13 \\
$N_{\rm buckled}$ & 15 & 25 & 38 & 23 & 0 & 0 \\
$f_{\rm bar}$ & $0.49^{+0.05}_{-0.06}$ & $0.82^{+0.05}_{-0.07}$ & $0.76^{+0.04}_{-0.05}$ & $0.72^{+0.04}_{-0.05}$ & $0.80^{+0.06}_{-0.08}$ & $0.76^{+0.09}_{-0.11}$ \\
\\
$f_{\rm buckled}$ & $0.42^{+0.08}_{-0.08}$ & $0.69^{+0.08}_{-0.08}$ & $0.55^{+0.06}_{-0.06}$ & $0.27^{+0.05}_{-0.05}$ & $0.00^{+0.04}_{-0.00}$ & $0.00^{+0.07}_{-0.00}$ \\
\\

\hline \hline

\\
 & $8<\log{M_*}<10$ & $10<\log{M_*}<10.5$ & $10.5<\log{M_*}<11$ &$11<\log{M_*}<11.5$ & $11.5<\log{M_*}<12$ & \\
\hline
\\
$N_{\rm disk}$ & 27 & 63 & 158 & 101 & 16 \\
$N_{\rm bar}$ & 22 & 46 & 119 & 60 & 9 \\
$N_{\rm buckled}$ & 0 & 3 & 51 & 40 & 7 \\
$f_{\rm bar}$ & $0.82^{+0.07}_{-0.08}$ & $0.73^{+0.05}_{-0.06}$ & $0.75^{+0.04}_{-0.03}$ & $0.59^{+0.05}_{-0.05}$ & $0.56^{+0.12}_{-0.12}$ &  \\
\\
$f_{\rm buckled}$ & $0.00^{+0.04}_{-0.00}$ & $0.07^{+0.04}_{-0.03}$ & $0.43^{+0.04}_{-0.05}$ & $0.67^{+0.05}_{-0.07}$ & $0.78^{+0.10}_{-0.16}$ & \\
\\
\hline \hline
\\
 & $0.6 < V-I < 1.0$ & $1.0 < V-I < 1.2$ & $1.2 < V-I < 1.4$ & $1.4 < V-I < 1.9$ & & \\
\hline
\\
$N_{\rm disk}$ & 52 & 135 & 147 & 33 \\
$N_{\rm bar}$ & 40 & 105 & 88 & 23 \\
$N_{\rm buckled}$ & 4 & 33 & 50 & 13 \\
$f_{\rm bar}$ & $0.77^{+0.05}_{-0.06}$ & $0.78^{+0.03}_{-0.04}$ & $0.60^{+0.04}_{-0.04}$ & $0.70^{+0.07}_{-0.09}$ & &  \\
\\
$f_{\rm buckled}$ & $0.10^{+0.06}_{-0.04}$ & $0.31^{+0.05}_{-0.04}$ & $0.57^{+0.05}_{-0.06}$ & $0.56^{+0.09}_{-0.11}$ & & \\
\\

\hline \hline
\\
 & $0 < \frac{M_{\rm H~I}}{M_{*}} < 0.02$ & $0.02 < \frac{M_{\rm H~I}}{M_{*}} < 0.05$ & $0.05 < \frac{M_{\rm H~I}}{M_{*}} < 0.07$ & $0.07 < \frac{M_{\rm H~I}}{M_{*}} < 0.1$ & $0.1 < \frac{M_{\rm H~I}}{M_{*}} < 0.5$ & \\
\hline
\\
$N_{\rm disk}$ & 104 & 117 & 62 & 36 & 45 \\
$N_{\rm bar}$ & 66 & 79 & 48 & 26 & 36 \\
$N_{\rm buckled}$ & 28 & 35 & 26 & 10 & 2 \\
$f_{\rm bar}$ & $0.63^{+0.05}_{-0.04}$ & $0.68^{+0.04}_{-0.05}$ & $0.77^{+0.05}_{-0.05}$ & $0.72^{+0.07}_{-0.08}$ & $0.80^{+0.05}_{-0.07}$ &  \\
\\
$f_{\rm buckled}$ & $0.42^{+0.07}_{-0.06}$ & $0.44^{+0.06}_{-0.05}$ & $0.54^{+0.07}_{-0.07}$ & $0.38^{+0.10}_{-0.08}$ & $0.06^{+0.05}_{-0.03}$ & \\

\enddata 
\tablecomments{The bar fraction ($f_{\rm bar} = N_{\rm bar} / N_{\rm disk}$), buckled bar fraction ($f_{\rm buckled} = N_{\rm buckled} / N_{\rm bar}$), B/PS fraction ($f_{\rm B/PS} = N_{\rm B/PS} / N_{\rm buckled}$) and barlens fraction ($f_{\rm barlens} = N_{\rm barlens} / N_{\rm buckled}$) in bins of galaxy inclination angle ($i$), morphology, stellar mass ($\log{M_{*}}$), color ($V-I$), and gas mass ratio ($M_{\rm H~I} / M_{*}$).} 
\label{tab:stat}
\end{deluxetable*}

\subsection{A possible buckling bar candidate: ESO 506$-$G004}

\begin{figure}
\epsscale{1.2}
\plotone{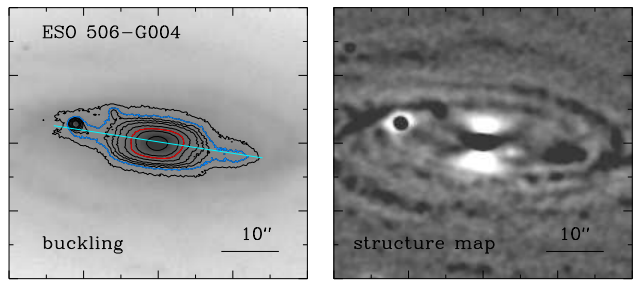}
\caption{A candidate buckling bar in ESO 506$-$G004. The left panel shows the CGS $I$-band image, and the right panel shows a structure map generated from the image. In the left panel, the red and blue contours show the inner boxy isophote and the extended spur morphology. The cyan solid line represents the major axis of the red boxy isophote. The two spurs are in the same side compared to the cyan line, which provides evidence of the buckling process. The X-shaped morphology can be seen from the structure map in the right panel.}
\epsscale{1.}
\label{fig:buckling}
\end{figure}

It is interesting to search for examples of peanut-shaped bulges that are currently experiencing the buckling process. Considering the short timescale for buckling to occur, only a small fraction of bars are expected to be undergoing this process at any given time. Based on numerical simulations, \citet{erw_deb_16} pointed out that bars in the process of buckling often show vertical asymmetry, in particular trapezoidal inner regions and outer spurs that are identically offset from the inner region major axis. Following these criteria, we visually examined the CGS barred disks again, and found one possible candidate, ESO 506$-$G004. The isophotes and the structure map\footnote{Similar to unsharp masking, the structure map developed by \citet{pog_mar_02} is particularly effective in enhancing spatial variations on the smallest resolvable scales, namely that of the point-spread function. It is effective in identifying X-shaped structures. The detailed calculation of structure maps for CGS is given in Paper I.} of this galaxy are shown in Figure~\ref{fig:buckling}. The inner region is slightly trapezoidal, with the outer spurs deviating in the same direction. 

The isophotal structure is similar to the buckling galaxy NGC~4569 reported by \citet{erw_deb_16}. Although the trapezoidal inner isophotes of ESO~506$-$G004 are not as prominent as those of NGC~4565, the offset of the outer spurs with respect to the inner region major axis is quite similar between the two galaxies. From the structure map we identify an X-shaped bar, an expected signature of the buckling process in the disk. According to numerical simulations, once a bar forms, it quickly buckles in the vertical direction (within $\sim0.5$ Gyr). This probably indicates that the bar in this galaxy formed fairly recently. The confirmation of the buckling bar in ESO~506$-$G004 would require stellar kinematic information. \citet{erw_deb_16} pointed out that the buckling phase is accompanied by asymmetries in the stellar velocity dispersion measured along the major axis of the bar.

In the sample of \citet{erw_deb_16}, for the 44 barred galaxies with $M_{*} \geq 10^{10.4} \, M_{\odot}$,  they found a B/PS fraction of $\sim$80\% and a buckling fraction of 4.5\%. Using a toy model, \citet{erw_deb_16} estimated that after bar formation, the time delay for buckling is $\sim$2.2 Gyr, and the buckling lasts $\sim$0.8 Gyr. The results are roughly consistent with simulations, which find that the time delay for buckling is $1-2$ Gyr, and that the buckling phase lasts $0.5-1$ Gyr \citep{mar_shl_04, martin_etal_06}. Our sample is independent from that of  \citet{erw_deb_16}, and we include both the barlens and B/PS bulges to trace the buckled bar in a more comprehensive way. In CGS, there are 195 barred disks with $M_{*} \geq 10^{10.4} \, M_{\odot}$. In this subsample, 98 are buckled, with 36 B/PS bulges and 62 barlenses; the corresponding buckled fraction is 50\% (18\% for B/PS bulges and 32\% for barlenses), and the buckling fraction is 0.5\%. These values are lower than those in \citet{erw_deb_16}, suggesting shorter buckling time delay and buckling phase period as given by these authors.  The buckling bars are difficult to identify in face-on systems. Thus, the buckling fraction we find here is likely a lower limit to the intrinsic buckling fraction. This difference may be due to several factors, such as sample selection and methods for identification of B/PS structures and barlenses. 


It is not likely that we missed a large fraction of B/PS bulges in the CGS sample. Although we use visual examination to identify barlenses and B/PS bulges, from Figure~\ref{fig:frac_inc}, the fraction of buckled bars identified is quite flat as a function of inclination ranging from $20^\degree$ to $70^\degree$, while the B/PS and barlens fractions each show a strong dependence on inclination angle. If our visual search had a much lower efficiency for identifying buckled bars at either high or low inclination angles, this would manifest itself as a strong dependence of $f_\mathrm{buckled}$ on $i$, contrary to what we observe.



\section{DISCUSSION}
\label{sec:discussion}

\subsection{Conditions for bar formation and buckling}

In this section we discuss the conditions for disks to grow bars and for the bars to buckle. Numerical simulations suggest that to support the growth of a bar perturbation the disk should be dynamically cold. Our analysis indicates that the bar fraction is relatively higher in bluer, lower mass late-type disks with high gas mass fraction (Fig.~\ref{fig:frac}).

\begin{figure}
\epsscale{1}
\plotone{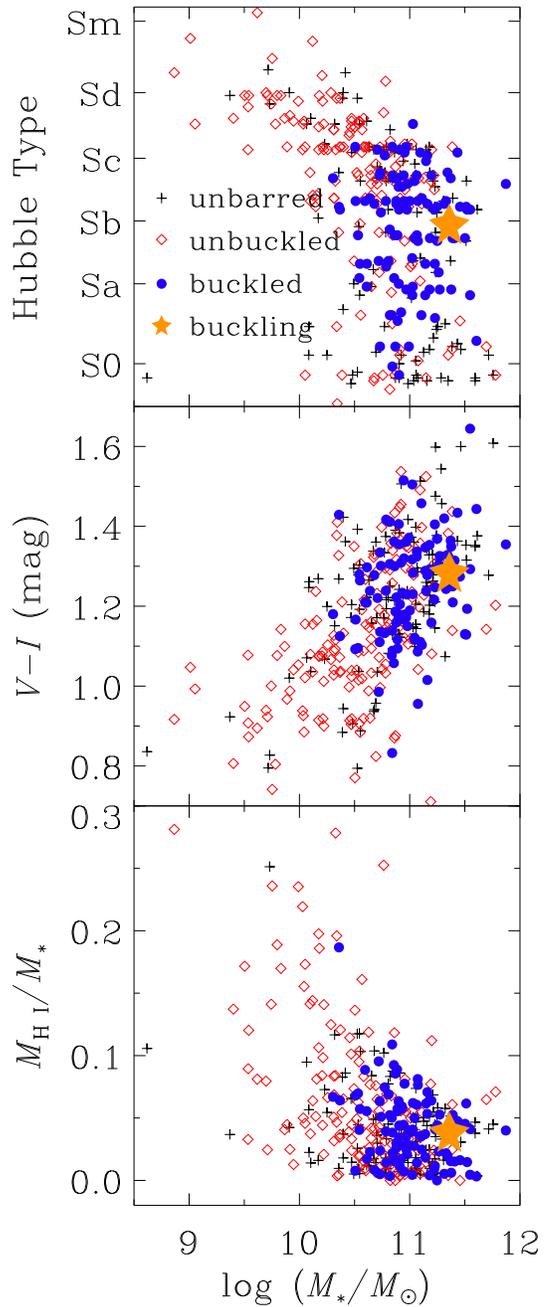}
\caption{Distributions Hubble type, color, and gas mass ratio as a function of stellar mass. Different symbols represent different types of galaxies (black cross: unbarred disks; red diamond: unbuckled bars; blue dots: buckled bars; brown star: buckling bar).}
\epsscale{1.}
\label{fig:corr}
\end{figure}

To understand the conditions for bar formation and buckling, it is essential to investigate the distribution of unbarred, unbuckled, and buckled disks as a function of galaxy physical parameters. Figure~\ref{fig:corr} shows the results of our sample. Comparing the unbarred disks with the barred disks (including both buckled and unbuckled bars), it seems that they show similar distributions. However, there are indeed more barred disks with lower stellar mass, higher gas mass ratio and bluer color. This is consistent with the expectations from numerical simulations that dynamically colder disks can better support the initial growth of the bar instability. In the local Universe, early-type galaxies usually have higher stellar mass than late-types. Based on the morphological classification, the early-type disks also harbor significant classical bulges compared to the late-type ones, where low-mass disk-like pseudobulges are seen more frequently. In addition, the higher fraction of gas mass in the late-type disks also helps to reduce the velocity dispersion, resulting in conditions more conducive to the formation of bars.

From Figure~\ref{fig:frac}, it can be seen that the buckled bar fraction is much higher in early-type and massive disks. It also increases toward galaxies with red colors and lower gas mass ratios. Among these parameters, the stellar mass and gas mass ratio are more important. As shown in  Figure~\ref{fig:frac_grid_buck} (e) and Figure~\ref{fig:corr}, most buckled bars occur in systems with $M_{*} > 10^{10.5} \, M_{\odot}$ and $M_{\rm H~I}/M_{*} < 0.1$. The buckling bar candidate ESO~506$-$G004 lies in the buckled bar region of this parameter space.

In simulations, once the bar forms, it easily gets buckled in the subsequent evolution stage. However, the observational results portray a more complicated picture. Based on our sample, it seems that for a bar to buckle, it needs to stay in a massive disk with less gas content. This is different from the bar formation condition, which primarily requires a dynamically cold disk. The buckling instability mainly perturbs the stellar orbits constituting the bar in the vertical direction, while still maintaining the orbital shapes in the disk plane. It happens when $\sigma_{Z} \leq 0.3\sigma_{R}$ \citep{toomre_64}. Once the bar forms, in the inner region of the galaxy, a massive disk with deeper potential well can better constrain the growth of the vertical velocity dispersion, compared to a low-mass disk. In addition, the bar also needs to be massive enough to maintain the orbits after their vertical resonances to form B/PS bulges and barlenses. From our results, it seems that the lower boundary of the disk stellar mass for a bar to buckle is $\sim 10^{10.5}\,  M_{\odot}$.

\subsection{Comparison with previous work}

\citet{erw_deb_17} found that the frequency of B/PS bulges strongly depends on galaxy stellar mass, with $\sim$80\% bars in galaxies more massive than $10^{10.4} \ M_{\odot}$ having B/PS bulges. They found strong trends of B/PS fraction with Hubble type, with a B/PS fraction of $\sim$80\% for S0--Sbc galaxies and 15\% for disk galaxies of later type. In addition, they reported no evidence that the gas mass ratio affects the presence of B/PS bulges.

Our results in this study are generally consistent with their findings. It is worth noting that we consider both B/PS bulges in moderately inclined disks and barlenses in face-on disks, which have been suggested to be the same structure seen from different viewing angles. \citet{erw_deb_17} only focused on the B/PS bulges in moderately inclined disks. Our sample has the additional advantages of being larger and more statistically complete than the sample in \citet{erw_deb_17}.

Strong trends with the Hubble type also exist in our results. The bar fraction in disks increases from S0 ($\sim$50\%) to Sd ($\sim$80\%) galaxies. But the buckled fraction in bars shows different trends. In S0s, the buckled bar fraction is about 30\%; it peaks in Sa and Sb galaxies, which have a buckled bar fraction of  $\sim$70\%. For later type galaxies, the buckled fraction decreases significantly. The results are similar to \citet{erw_deb_17} except that they found a B/PS fraction about 80\% in S0s.

We also find that the buckled bar fraction increases quickly for stellar masses larger than $10^{10.5} \ M_{\odot}$, rising quickly  from 0\% to $\sim$80\%  beyond this threshold. By contrast, the bar fraction gradually drops from 80\% to 50\% as the stellar mass increases. The stellar mass seems to play different roles here. On the one hand, it significantly enhances bar buckling. On the other hand, bar formation seems to be restrained at the high-mass end. 

In our study, gas also appears to play an important role. The bar fraction slightly increases from 60\% to 80\% at higher gas mass fraction; however, the buckled bar fraction drops significantly at high gas mass ratio. This is consistent with simulation predictions. 

The size distribution of B/PS bulges was investigated in \citet{erw_deb_17}. Generally speaking, the B/PS bulge extends to about half the bar length, which is consistent with numerical simulations. Here we only focus on general statistical properties of the galaxies. A future work will investigate the geometric parameters (e.g., ellipticity, size, etc.) and isophotal parameters (e.g., surface brightness profile, light fraction, etc.) of the B/PS, barlens, and unbuckled bars.

\section{SUMMARY}
\label{sec:summary}

We perform a statistical study of the host galaxy properties of bars and buckled bars in a large, statistically complete sample of local disk galaxies from CGS.  With careful visual examination, B/PS bulges and barlenses are identified. These structures constitute the main features of a buckled bar seen from different viewing angles. Across different inclination angles, the bar fraction and buckled bar fraction are roughly constant, indicating the robustness of the bar/buckled bar detection.

We investigate the statistical incidence of bars and buckled bars as a function of host galaxy Hubble type, stellar mass, color, and gas mass ratio. The bar fraction increases from 50\% in S0s to $\sim$80\% in later types. It decreases toward massive stellar disks and increases with higher gas mass ratios. From the distribution of these parameters, we find no clear differences between bars and unbarred disks. It seems quite difficult to prevent bar formation. On the other hand, the buckled bar fraction increases significantly toward systems of high stellar mass ($>60\%$ with $M_{*} > 10^{10.5}\,  M_{\odot}$) and low gas mass ratio ($M_{\rm H~I} / M_{*} < 0.1$). 

Overall,  stellar mass is one of the most important parameters examined, but gas mass fraction also appears to be related to bar properties. Bar formation requires a dynamically cold disk, where a massive disk with a classical bulge may prevent bar formation in the first place. However, once the bar forms, a deeper potential well seems necessary to support the buckled bar orbits. 

We also report a buckling bar candidate in ESO~506$-$G004.  Since the buckling stage is usually very short, it is rare to see such an ongoing event. Kinematic observations of this galaxy may provide further tests to determine whether this bar exhibits signs of buckling.

We thank the anonymous referee for constructive suggestions that helped to improve the paper. ZYL is supported by a National Natural Science Foundation of China grant (11403072), a China-Chile joint grant from CASSACA, a Shanghai Yangfan Research Grant (14YF1407700), and a LAMOST Fellowship, which is supported by Special Funding for Advanced Users, budgeted and administered by the Center for Astronomical Mega-Science, Chinese Academy of Sciences (CAMS). LCH is supported by the National Key Program for Science and Technology Research and Development (2016YFA0400702) and the National Science Foundation of China (11303008, 11473002). This work made use of the facilities of the Center for High Performance Computing at Shanghai Astronomical Observatory. 


\end{document}